\documentclass[twocolumn, amssymb,amsmath, aps, prb, showpacs, 10pt]{revtex4-2}
\usepackage{graphicx}
\usepackage{color}
\usepackage{multirow}
\usepackage{hyperref} 
\usepackage{amsmath}
\usepackage{amssymb}
\usepackage{epsfig}
\usepackage{steinmetz}
\DeclareMathAlphabet{\pazocal}{OMS}{zplm}{m}{n}
\usepackage{graphicx}
\usepackage{color}
\usepackage{bm}
\usepackage[normalem]{ulem}
\usepackage{soul}
\usepackage{braket}
\usepackage{csquotes}

\begin{document}
	\title{Observation of Griffiths-like phase in the quaternary Heusler compound NiFeTiSn}
\author{Snehashish Chatterjee, Saurav Giri, Subham Majumdar}
\email{sspsm2@iacs.res.in}
\affiliation{School of Physical Sciences, Indian Association for the Cultivation of Science, 2A \& B Raja S. C. Mullick Road, Jadavpur, Kolkata 700032, India}
\author{Prabir Dutta}
\affiliation{Jawaharlal Nehru Centre for Advanced Scientific Research,Jakkur, Bangalore, 560064, India}
\author{Pintu Singha,Aritra Banerjee}
\affiliation{Department of Physics, University of Calcutta, Kolkata 700 009, INDIA}
\begin{abstract}
  The quaternary Heusler compound NiFeTiSn can be considered to be derived from the exotic pseudogap-compound Fe$_2$TiSn by the replacement of one Fe atom by Ni. In contrast to Fe$_2$TiSn, which shows a disorder induced ferromagnetic phase,  the ground state of NiFeTiSn is antiferromagnetic with the signature of spin canting. Interestingly, NiFeTiSn shows a Griffiths-like phase characterized by isolated ferromagnetic clusters before attaining the antiferromagnetic state. The Griffiths-like phase is possibly associated with the antisite disorder between Fe and Ti sites as evident from our powder X-ray diffraction study.  The compound also shows rather unusual temperature dependence of resistivity, which can be accounted by the prevailing structural disorder in the system. NiFeTiSn turned out to be a rare example where Griffiths-like phase is observed in a semiconducting  3$d$ transition metal based intermetallic compound with antiferromagnetic ground state.            
\end{abstract}
\maketitle

\section{Introduction}
In recent times, transition metal based Heusler compounds and alloys~\cite{heusler} are in the limelight of active research by virtue of their diverse and novel electronic properties, which include half-metallic ferromagnetism, spin-gapless semiconductivity, Weyl semimetallicity, magnetic shape memory effect, high Seebeck effect, magneto- and baro-caloric effects, unconventional superconductivity, anomalous Hall effect and so on~\cite{parkin,kondo,thermo}. Among others,  Fe$_2$TiSn deserves a special mention in the Heusler family due to its fascinating electronic and magnetic properties. First principle density functional theory (DFT) based calculation indicates the compound to have a nonmagnetic ground state with a pseudogap at the Fermi level ($E_F$)~\cite{fe2tisn}. The number of valence electrons ($Z_t$) is 24 in Fe$_2$TiSn, which implies zero moment according to the Slater-Pauling rule~\cite{sp}. However, actual experimental measurement indicates that the compound is a weak ferromagnet with Curie temperature ($T_C$) close to 260 K and saturation moment 0.2 $\mu_B$/f.u.~\cite{fe2tisn1}. The possible origin of ferromagnetism is ascribed to the antisite disorder between Fe and Ti sites, a phenomenon which is very common among Heusler compositions. Initial analysis of the low-temperature  heat capacity and the resistivity data led to the conclusion that the compound was a heavy fermion~\cite{fe2tisn}. Optical conductivity measurements, however, shows a simple Drude-like response of free carriers ruling out the Kondo or heavy fermion like state~\cite{fe2tisn, fe2tisnoptical}. The low temperature anomaly in Fe$_2$TiSn is believed to arise from clustering of atoms, which is also in accordance with the picture of disorder induced ferromagetism in the system. Notably, the compound also shows large positive Seebeck coefficient at room temperature.

\par
Fully ordered Fe$_2$TiSn crystallizes in the cubic L2$_1$ structure with space group $Fm\bar{3}m$, where Fe atoms occupy 8c (1/4, 1/4, 1/4) Wyckoff position. In case of an L2$_1$ full Heusler compound with general formula X$_2$YZ, the replacement of one X atom with a different transition metal (X$^{\prime}$) leads to a change in structural symmetry. The resultant compound (XX$^{\prime}$YZ) retains the cubic symmetry but attains a different space group, $F\bar{4}3m$. The study of such  quaternary Heusler compounds have accelerated drastically in last two decades because of their possible application as a spin-polarized material in spintronics (CoFeMnGa, CoFeMnGe, CoMnCrAl)~\cite{cofetisn,alijani,suresh,enamul}. Apart from their half metallic character, some of the quaternary compounds, such as  CoFeMnSi, CoFeCrGa ~\cite{cofemnsi,cofecrga}, are found to be  spin-gapless-semiconductors.

\par       
Considering rich physics associated with the full Heusler Fe$_2$TiSn, it might be tempting to study the quaternary Heusler compound derived from it. There are previous reports on doping at the Fe site, however, a detailed study on the quaternary derivative is lacking. It is already known that Ni$_2$TiSn is a Pauli paramagnet with vanishing local moment at the Ni site~\cite{wrona}. Therefore, it is worth studying the quaternary compound NiFeTiSn, where one Fe site is diluted by a Ni atom. Such substitution will not only reduce the Fe concentration, it will also change the local crystallographic environment of Fe due to the change in space group.    

\par
The degree of antisite disorder depends on the atomic radii of the constituent metals, and it is likely to differ in case of NiFeTiSn. In the previous work on Fe$_{2-x}$M$_x$TiSn (M = Ni, Co), superparamagnetic-like state has been predicted~\cite{maple}. Also, doping is found to destroy the pseudogap at the $E_F$. Our present study on NiFeTiSn indicates an antiferromagnetic (AFM)-like state at very low temperature (below about 7 K), which is in sharp contrast with the FM state reported for Fe$_2$TiSn. None the less, NiFeTiSn shows a non-Curie-Weiss-like state below about 275 K indicating the presence of short range magnetic correlations. Our investigations suggest that this state with short range magnetic correlations complies with the properties of the Griffiths phase (GP). In GP, the magnetic correlation does not vanish immediately above the long range ordering temperature, but it continues to exist up to a higher temperature~\cite{griffiths1,griffiths2,griffiths3,griffiths4,giri}. In systems having ordered magnetic ground state, GP  occurs due to the percolative nature of the distribution of short-range ferromagnetic (FM) clusters developed  much above the  ordering temperature in the magnetically disordered paramagnetic (PM) regime.       
\begin{figure}
	\centering
	\includegraphics[width = 8 cm]{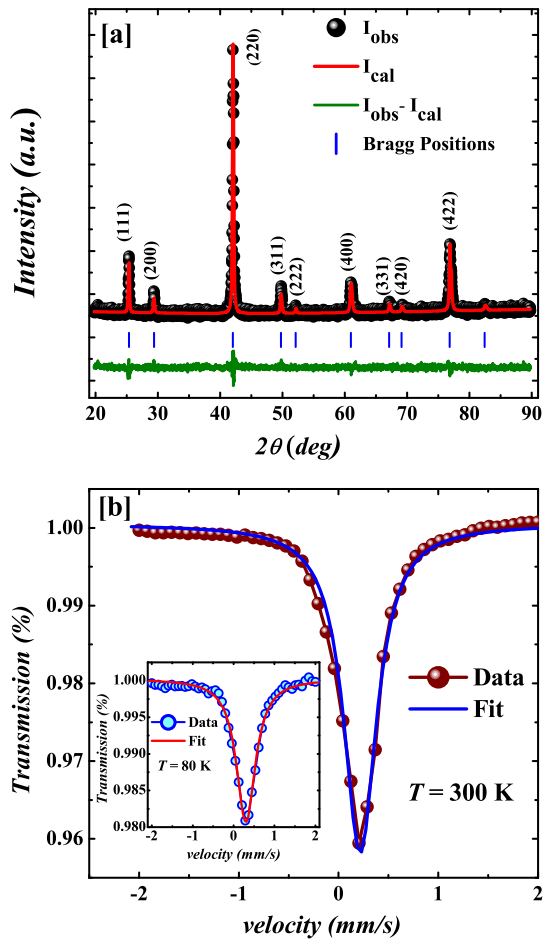}
	\vskip 0.5 cm
	\caption {(a) shows the powder X-ray diffraction pattern of NiFeTiSn (data points) at room temperature using Cu K$_{\alpha}$ radiation. The solid line is the Rietveld refinement fit to the data. (b) shows the  M\"ossbauer spectrum of NiFeTiSn recorded at room temperature. The 80 K spectrum is shown in the inset.}
	\label{xrd}
\end{figure}
\par
Magnetic materials, in presence of certain degree of disorder, can give rise to this remarkable GP-like  magnetic state. In majority of the cases, the GP is followed by a long range FM order at low temperature. The occurrence of GP in the AFM compounds are mostly reported in pnictides, manganites, perovskites~\cite{fecos,pramanik,LaFeMnO,PrCoMnO,arindam1,arindam2} and few other rare-earth intermetallic alloys~\cite{gdsn2,GP3,GdGe}. On the contrary, NiFeTiSn is a transition metal based intermetallic system with  rather simple cubic structure. Our careful investigation and analyses indicate that the ground state of NiFeTiSn can better be described by a canted AFM one. In addition, the compound shows negative temperature coefficient of resistivity  with rather unusual temperature dependence.

\section{Experimental Details}
Polycrystalline  NiFeTiSn was prepared using the stoichiometric ratio  of the constituent elements by arc melting procedure in an inert argon atmosphere followed by annealing at 800$^{\circ}$ C for 3 days. Powder X-ray  diffraction (PXRD) was performed at room temperature and the crystallographic structure was analysed using the Rietveld refinement technique with MAUD software package~\cite{maud}.  Magnetic measurements were carried out with the help of  a commercial Quantum Design SQUID-VSM. The resistivity ($\rho$) was measured on a cryogen-free high magnetic field system (Cryogenic Ltd., U.K.) between 5-300 K. Thermoelectric measurements were carried out in the temperature range of 10 K-300 K using the differential technique. Heat capacity was measured using a Quantum Design physical properties measurement system. 
\par
To confirm the chemical composition of the sample, we performed a semi-quantitative Energy Dispersive X-Ray (EDX) analysis over multiple areas of the samples, and the average composition from  the EDX analysis is found to be Ni$_{25.1}$Fe$_{23.6}$Ti$_{24.8}$Sn$_{26.6}$. From the obtained value, the equiatomic quaternary compound can be considered to be stoichiometric within the accuracy (2-5\%~\cite{edx}) of the EDX technique. $^{\rm 57}$Fe M\"ossbauer spectra were recorded using an alternating constant WissEl M\"ossbauer spectrometer. The system operates in a horizontal transmission geometry with source, absorber and detector in a linear arrangement.  Isomer shifts were measured with reference to $\alpha$-iron metal foil at the same temperatures. Normos-A M\"ossbauer Fit program was used to fit the experimental data.   
\begin{figure}
	\centering
	\includegraphics[width = 8 cm]{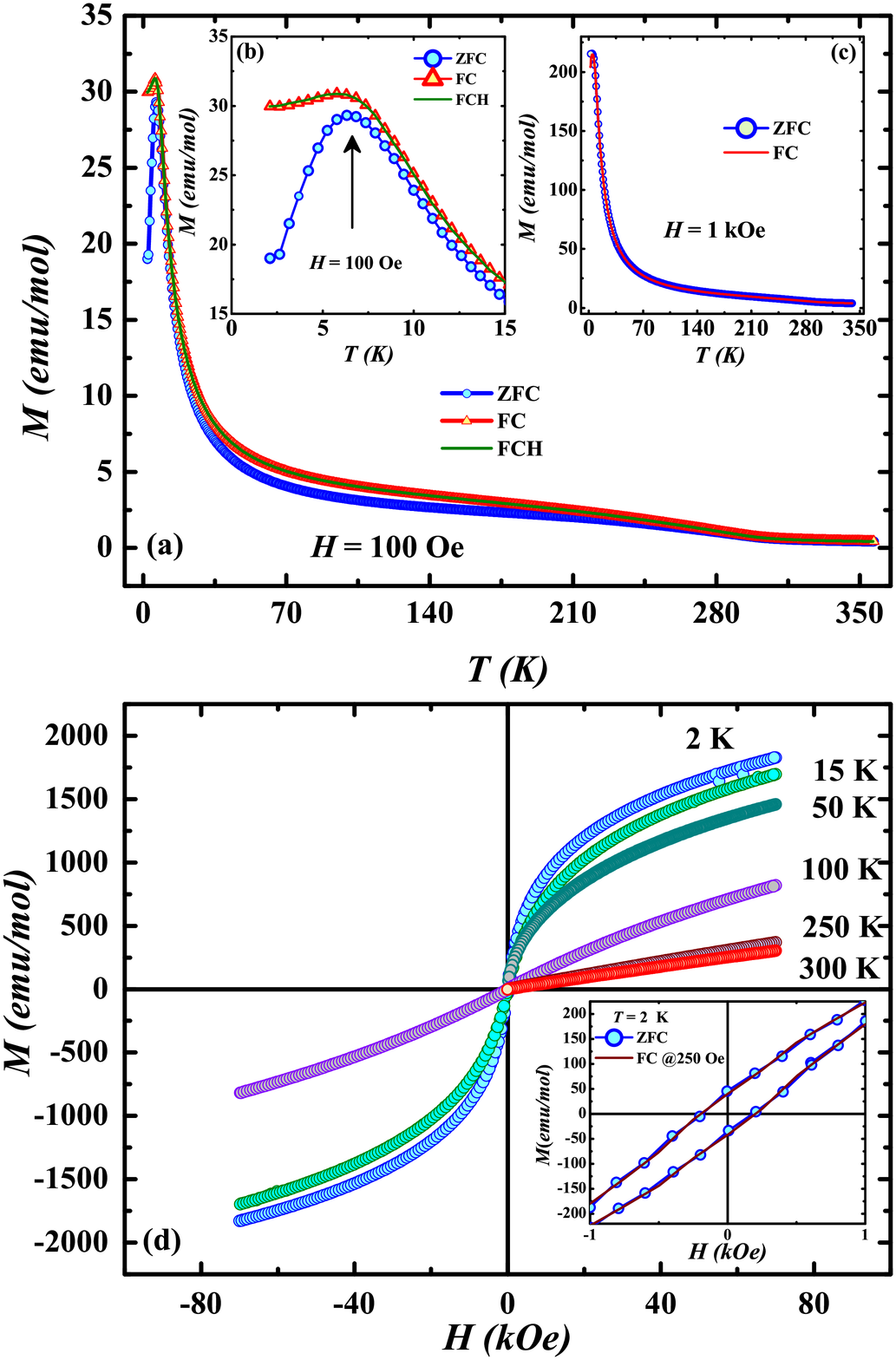}
	\caption {(a) Magnetisation versus temperature plots  in zero-field-cooled (ZFC), field-cooled (FC) and field-cooled-heating (FCH) protocols for NiFeTiSn measured under 100 Oe of magnetic field. (b) shows an enlarged view of the low temperature data highlighting the antiferromagnetic transition. (c) shows $M(T)$ for $H$ = 1 kOe. (d) shows the isothermal magnetisation curves recorded at different constant temperatures. Inset shows an enlarged view of $M(H)$ at $T$ = 2 K recorded in ZFC and FC condition.}
	\label{mag}
\end{figure}

\section{Results}
\subsection{Powder X-ray Diffraction}   
Room temperature powder X-ray diffraction pattern for the compound NiFeTiSn  is shown in Fig~\ref{xrd} (a). From the observed  reflections and subsequent refinements, the sample is found to crystallize in the cubic Y-II type structure without any impurity phases. In this structure, Ni, Fe, Ti and Sn occupy the lattice sites 4$b$(1/2, 1/2, 1/2), 4$d$(3/4, 3/4, 3/4), 4$c$(1/4, 1/4, 1/4) and 4$a$ (0,0,0) respectively~\cite{gao,rasool}. Both the superlattice reflection (111) and (200) are found to be present, indicating a predominantly ordered structure. However, for an ideal Y-type structure, the intensity of (111) and (200) in general remains the same. The reduction of (200) peak intensity in comparison to (111) peak ($I_{111}/I_{200} \sim$ 1.7) might be due to the existence of antisite disorder in the studied sample. The cubic lattice parameter  obtained from the  refinement is found to be 6.070 \AA~.
\par
It is important to obtain a quantitative idea of the degree of disorder from the intensities of the diffraction peaks. A rough estimation of the degree of site ordering of the constituent atoms can be made  by calculating the quantities $p = I_{200}/I_{220}$ and $q= I_{111}/I_{220}$, where $I_{hkl}$ is the intensity of the ($hkl$) line in the PXRD pattern. The order parameters, $\sigma$ and $\alpha$ are given by,  $\sigma^2 = q_{obs}/q_{fit}$ and $\sigma^2(1-2\alpha)^2 = p_{obs}/p_{fit}$~\cite{suresh}. Here the subscripts {\it obs} and {\it fit} correspond to the experimentally observed  and the fitted values respectively. The fitted intensities are obtained from the Reitveld refinement without considering any site disorder. For an ideally ordered system, $\sigma$ = 1 and $\alpha$ = 0, and for fully B2 disordered case, $\sigma$ = 1 and $\alpha$ = 0.5. We get, $\sigma$ = 1.08 and $\alpha$ = 0.1 for NiFeTiSn. This indicates the presence of disorder (Fe-Ti antisite disorder) in case of NiFeTiSn.     
\begin{figure}
	\centering
	\includegraphics[width = 8 cm]{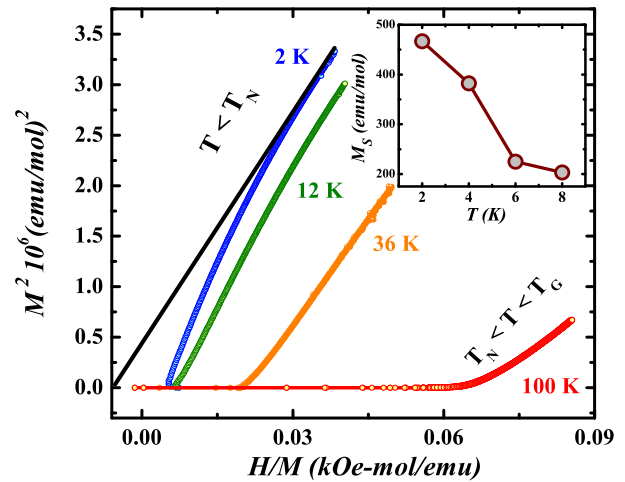}
	\vskip 0.5 cm
	\caption {Main panel presents the Arrot plot ($M^2$ vs $H/M$ ) of the sample recorded at different temperatures. Inset: Variation of spontaneous magnetisation ($M_S$) with $T$ below $T_N$.}
	\label{arrot}
\end{figure}


\par
M\"ossbauer spectra for NiFeTiSn [see fig~\ref{xrd}(b)] recorded at both 300 K and 80 K exhibit singlet pattern, which indicates that local electric field at Fe site in the compound is symmetric. These singlet M\"ossbauer peaks at 300 and 80 K are well fitted with single Lorentzian function~\cite{mossbauer1}. The Isomer shift is also found to be slightly higher at low temperature [0.24 mm/s (300 K) and 0.28 mm/s (80 K)].

\begin{figure}
	\centering
	\includegraphics[width = 8 cm]{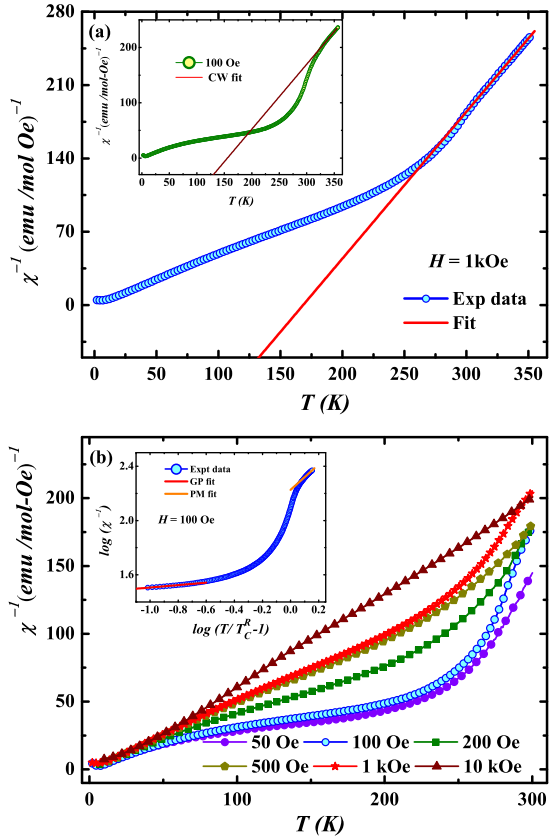}
	\caption {(a) Inverse susceptibility versus temperature data recorded under $H$ = 1 kOe. The inset shows the same data measured under 100 Oe. (b) shows inverse susceptibility data measured in different values of the applied field. Inset  shows the $\log$-$\log$ plot of inverse susceptibility as a function of ($T/T_{0}^R$ -1).}
	\label{invchi}
\end{figure}
\begin{figure}
\centering
\includegraphics[width = 8 cm]{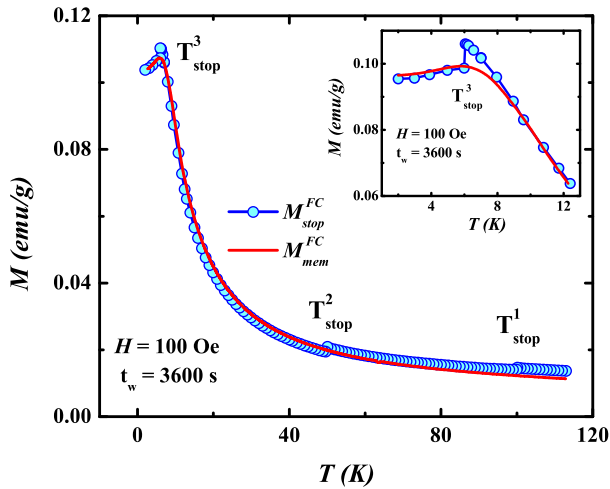}
\vskip 1 cm
\caption {Field-cooled-field-stop memory curves at three stopping temperatures, $T_{stop}$ = 6, 50 and 100 K. The inset shows an enlarged view of the low temperature data.}
\label{mem}
\end{figure}


\subsection{Magnetisation}
Temperature ($T$) variation of magnetisation ($M$) recorded for zero-field-cooled (ZFC), field-cooled (FC) and field-cooled heating (FCH) conditions under the magnetic fields, $H$ = 100 Oe and $H$ = 1 kOe are shown in fig.~\ref{mag}(a),(c). Both ZFC and FC data  are  characterized by a  peak around 7 K indicating an AFM-like transition (see  fig.~\ref{mag}(b)). More accurate value of the N\'eel temperature is obtained from the minimum of $dM/dT$ versus $T$ data, which is $T_N$ = 11 K. ZFC and FC curves are found to bifurcate from each other from around T$_{irr}$ = 270 K, which is much higher than $T_N$. There is a broad shoulder occurring around 250 K, and on further cooling, $M$ increases with $T$. This thermo-magnetic irreversibility well above $T_N$ and the presence of broad shoulder-like feature hint towards the existence of the short-range magnetic correlations in the backdrop of a PM phase~\cite{zhang}. Under $H$ = 1 kOe, the low-$T$ bifurcation between ZFC and FC curves becomes weak and shifts to lower-$T$ and the $T_{irr}$ almost vanishes.

\begin{figure}
	\centering
	\includegraphics[width = 8 cm]{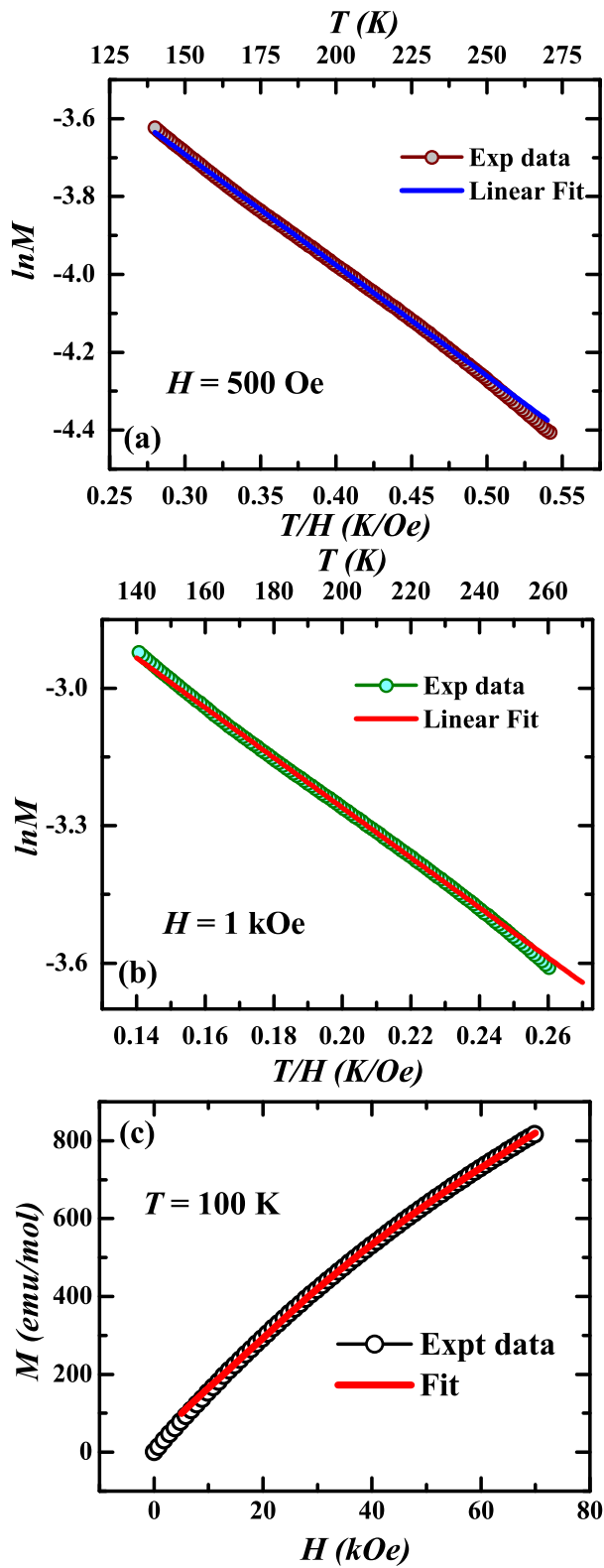}
	\caption {(a) and (b) shows $\log M$ versus $T/H$ plot of NiFeTiSn under the applied field of $H$= 500 Oe and 1 kOe. The solid line is the best linear fit using the relation~\ref{logMT}. (c) shows Magnetisation as a function of field at $T$ = 100 K following a power-law behaviour.} 
	\label{logM}
\end{figure}

\par
Fig.~\ref{mag}(d) shows the isothermal $M$ versus $H$ curves recorded at different constant temperatures in the ZFC condition ($T$ = 2, 15, 100, 250 and 300 K), between $H = \pm$ 70 kOe. The isotherms below 250 K follow an `S'-shaped nature with no saturation tendency even at $H$ = 70 kOe. The non-linear nature of the  $M(H)$ curves at temperatures well above $T_N$ [see 100 and 50 K isotherms in fig.~\ref{mag}(d)] reaffirms the existence of the short-range correlations. Inset of fig.~\ref{mag}(d) shows an enlarged view of $M(H)$ at $T$ = 2 K recorded in ZFC and FC conditions under $H$ = 250 Oe, indicating small coercive field close to 200 Oe and the absence of any exchange-bias in the compound.
 
 \par
Figure~\ref{arrot} shows the Arrot  ($M^2$ versus $H/M$) plot at few constant temperatures~\cite{arrot}. The curves are quasi-linear at higher values of $H/M$ for all temperatures above and below $T_N$. The linear extrapolation of the high-$H$ data shows positive intercept on the $M^2$ axis for the isotherms recorded below $T_N$. Such positive values of $M^2$ (below $T_N$), indicates the presence of spontaneous magnetisation, $M_S$ (as obtained from the square root of $M^2$). However, such positive intercept is absent  for the data recorded above $T_N$. This indicates that there exists a finite $M_S$  only below $T_N$, and  possibly the system has a canted antiferromagnetic ground state.

\par
In the main panel of  fig.~\ref{invchi}(a), we have plotted inverse susceptibility ($\chi^{-1} = H/M$ ) as a function of $T$. A linear region is observed between 310-360 K, and we have fitted this part using the Curie-Weiss (CW) law,  $\chi = C/(T-\theta_p)$, where $C$ is the Curie constant, $\theta_p$ is the paramagnetic Curie-Weiss temperature.  From the fitting, we obtain the effective paramagnetic moment, $\mu_{eff}$ to be  2.48 $\mu_B$, and  $\theta_p$ = 128 K. The observed large positive value of $\theta_p$ indicates the presence of dominant  FM correlations in the system. 
\par
 To shed some light on the observed behaviour, we have plotted $\chi^{-1}$ versus $T$ data over a wide range of temperatures. A strong downward deviation from the CW  law is observed from below about 300 K for the  100 Oe data (inset of fig.~\ref{invchi}(a)). This downward behaviour of $\chi^{-1}$ decreases with increasing $H$ and disappears for $H$ = 10 kOe [fig.~\ref{invchi}(b)].  The observation of a downward deviation in the $\chi^{-1}$ versus $T$ curve at temperatures much higher than magnetic ordering temperature (in this case $T_N$) is a hallmark of the presence of the Griffiths-like phase~\cite{zhang,eremina,nair}. The GP is characterized by the appearance of the finite size FM clusters in the backdrop of a PM phase well above the long range order  sets in~\cite{GP1,GP2}.  $\chi$ of the system fails to follow the typical Curie-Weiss law above $T_{N}$ up to a certain critical temperature known as Griffiths temperature ($T_{GP}$).  However, with increasing $H$, the downturn in the $\chi^{-1}$ vs $T$ data  softens and becomes gradually indistinguishable from the high-$T$ paramagnetic regime~\cite{GP3,GP4}. The application of a sufficiently high $H$ polarizes the spin on the outside part of the  clusters, leading to the suppression of the signature of  GP  in the $\chi^{-1}$ vs $T$ data~\cite{zhang,GP3}. The cluster-like FM component is masked at higher-$H$ by the linear increase of the PM contribution of the matrix~\cite{GP2}.

It is known that the inverse susceptibility in the  GP obeys a power law behaviour characterized by an exponent ($\lambda$)~\cite{griffiths1,pramanik},
\begin{equation}
 \chi^{-1} \propto (T-T_{0}^R)^{(1-\lambda)}
\label{griffithsfit}
\end{equation}
where, 0 $< \lambda <$ 1 for $T_{N} < T < T_{GP}$. In the PM phase, $\lambda \rightarrow$ 0, and eqn.~\ref{griffithsfit} turns into usual CW law. $T_{0}^R$ can be obtained from the paramagnetic phase by  setting $\lambda$ $\sim$ 0, which is close to the PM Curie temperature. We have plotted $\log$ ($\chi^{-1}$) against $\log$ ($T/T_{0}^R-1$) [inset of fig.~\ref{invchi}(b)] and the value of $\lambda$ is obtained from the slope of the linear fit using equation~\ref{griffithsfit} (solid red line fit). In PM and GP regions, the obtained values of $\lambda$ are 0.006 and 0.91 respectively (using $T_{0}^R$ = 147 K). Similar $\lambda$ value is earlier reported for  Heusler alloys Fe$_{2-x}$Mn$_x$CrAl showing Griffiths-like behaviour~\cite{kavita}. A large difference between T$_N$ and $T_{GP}$ is observed in  NiFeTiSn along with a value of $\lambda$ close to unity in the GP regime. This  signifies the strong deviation from the CW behaviour and point towards the robustness of the GP in this material. The deviation of  $\chi^{-1}(T)$ from CW law  prevails even  for $H$ as high as 10 kOe. This feature is also unusual compared to other systems where such  deviation vanishes for $H \gtrapprox$ 1 kOe~\cite{pramanik,fecos}.
\begin{figure}
	\centering
	\includegraphics[width = 8 cm]{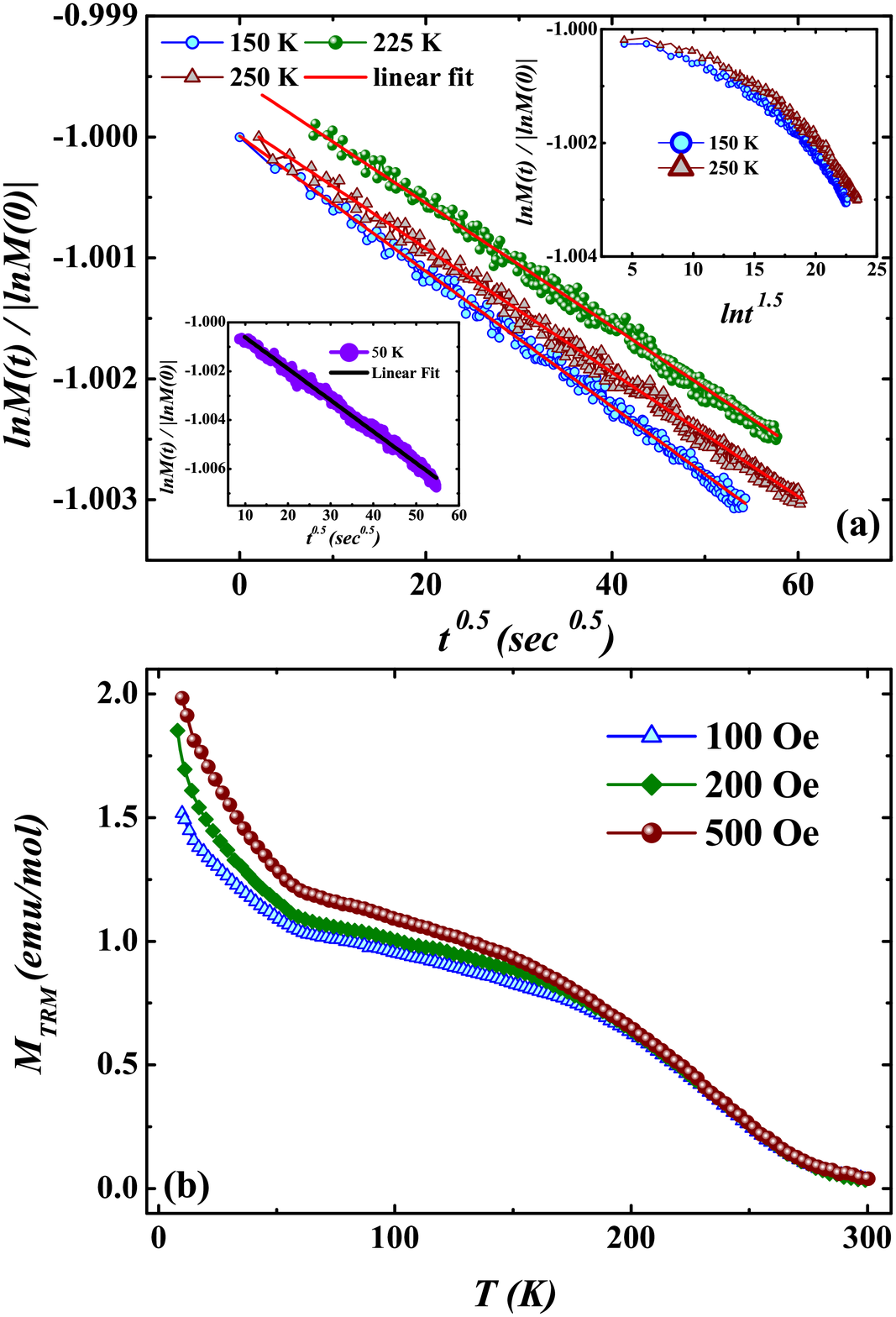}
	\caption {(a) shows thermo-remanent magnetisation after the sample being cooled in different fields. (b) Magnetic relaxation measured in the field cooled state at different constant temperatures, where $\log M$ is plotted as a function of $t^{0.5}$. We have also plotted $\log M$ as function of ($\log t^{1.5}$) (upper inset of (b)) }
	\label{relax}
\end{figure}
\par
 Galitski $\it{et. al.}$~\cite{sarma} studied the GP in the dilute magnetic semiconductor in the light of magnetic polarons. The magnetisation in the GP is found to be weak and it follows a relation:
\begin{equation}
 M(H,T) \propto\exp\left [-B\frac{T}{H}\right ]
\label{logMT}
\end{equation}
where, $B$ is a constant. It is to be noted that eqn.~\ref{logMT} does not take into account the pinning effect between the magnetic clusters which is relevant in the low field regime. Hence, the $M(T)$ data measured at lower fields ($H$ = 50, 100 and 200 Oe) cannot be fitted well using eqn~\ref{logMT}~\cite{lee}. In figure~\ref{logM}(a,b) we have plotted $\log M$ versus $T/H$ along with linear fit for $H$ = 500 Oe and 1 kOe within the temperature range 160 - 220 K. The value of the constant $B$ obtained from fitting is 28.4(3) and 54.5(4) Oe/K for $H$ = 500 Oe and 1 kOe respectively, which are  similar to the earlier reported values for other GP systems.~\cite{lee,nag}. 
\par
To confirm  GP in NiFeTiSn, 100 K $M(H)$ isotherm has been plotted  in figure~\ref{logM}(c). $M$ is found to obey the predicted GP behaviour given by, 

\begin{equation}	
M(H) = M_0 + aH^{\eta}
\label{hpower}
\end{equation}
	
where $M_0$ is the non-zero spontaneous magnetisation and $\eta$ is an exponent~\cite{blundell,neto}. The best fit between 20 to 70 kOe yields the exponent $\eta$ to be 0.78. This value of $\eta$ is close to the value of  $\lambda$ obtained from eqn.~\ref{griffithsfit}, and such closeness of these two exponents is common in GP~\cite{naka}.  Most importantly, a power law variation of $M(H)$ curve (eqn.~\ref{hpower}) is a clear deviation  from the linear $M(H)$ behaviour expected in a pure PM region (CW law)~\cite{blundell,kassis}. 

\par
To get further idea of the nature of the magnetic state in NiFeTiSn, we performed field-cooled-field-stop memory (FCFS memory) measurement as shown in fig.~\ref{mem}~\cite{memory1,memory2}. Initially, the sample was cooled under $H$ = 100 Oe from 300 to 2 K with intermediate stops at $T^{i}_{stop}$ = 6, 50 and 100 K  for $t_w$= 3600 s each.  The magnetic field was turned off at $T^{i}_{stop}$ which results in a step-like magnetization curve ($M^{FC}_{stop}$). After reaching 2 K, the sample was further heated in  presence of $H$ = 100 Oe  without any stops ($M^{FC}_{mem}$) [see fig.~\ref{mem}]. Interestingly, the heating  curve, $M^{FC}_{mem}$, does not show any anomaly at the stopping temperatures where the sample was allowed to stop during cooling. This points to the fact that no FC memory is present in the system, ruling out the possibility of a spin-glass or superparamagnetic state. 

\par
We  further studied  the  thermoremanant magnetisation (TRM), $M_{TRM}$, to characterize the GP ~\cite{trm,nair}. The protocol involves cooling the  sample from well above the magnetic transition temperature in the presence of cooling  field, $H_{cool}$. The field is  removed below $T_{N}$, and the magnetisation is measured while heating under $H$ = 0. In the present case, this protocol was repeated for three different cooling fields, $H_{cool}$ = 100, 200 and 500 Oe [fig~\ref{relax}(a).] The zero-field measurements  have the advantage that the contributions from the paramagnetic susceptibility are suppressed compared to an in–field measurement. All the three curves (recorded in zero field while heating  after being cooled in different values of  $H_{cool}$) show sizeable TRM below 300 K.  At temperature above about 210 K, these  curves are practically indistinguishable. Below about 200 K, the curves start to  diverge from each other and a hump-like feature is observed.  On field cooling, the FM clusters of the GP  are created  favourably  with the direction of $H_{cool}$. When the sample is heated from the lowest temperature after setting $H$ = 0, the clusters remains pinned and provide us a remanent magnetisation. The TRM eventually disappears above a certain temperature when the GP is destroyed. The presence of TRM supports the existence of magnetic clusters in the material. 
\begin{figure}
	\centering
	\includegraphics[width =  8 cm]{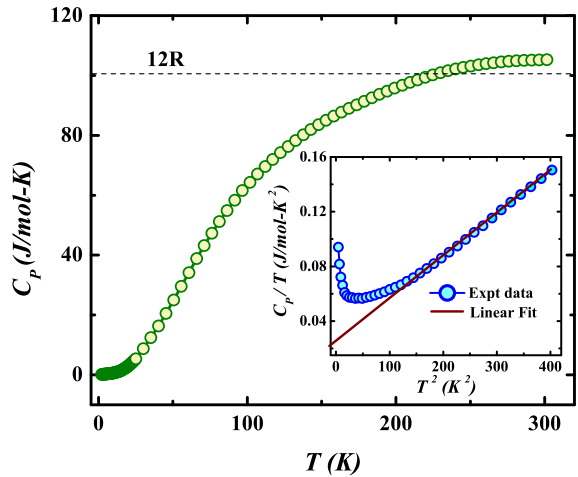}
	\caption {Heat capacity as a function of temperature. The inset shows the low-$T$ $C_p$ versus $T^2$ plot and solid line is a linear fit to the data.}
	\label{cp}
\end{figure}
\par
The short-range FM regions in the GP drastically influence the dynamics of the magnetic susceptibility. In the GP, the time required to reverse the effective spin of  clusters would be larger, resulting in a non-exponential decay of the spin-auto correlation function~\cite{GPmemory1}. For randomly diluted ferromagnets, this decay has been calculated to be of the form $\sim\exp(-Bt^{0.5}$)  and $\sim$ $\exp[-A\ln t^{(1.5)}]$  for 3D Heisenberg and Ising spin systems respectively~\cite{GPmemory2}. Fig.~\ref{relax}(b) highlights the isothermal remanent magnetisation (IRM) data for NiFeTiSn within the GP regime (50, 150, 225, and 250 K). We cooled the sample from 300 K in presence of $H$ = 50 Oe to a particular temperature of measurement, subsequently the time variation of $M$ was recorded  for 3600 s after switching off the magnetic field. Good linear fit is obtained for $\log M$ versus $t^{0.5}$ data for all the temperatures. The upper inset of fig.~\ref{relax}(b) presents the plot for $\log M$ versus $\log t^{1.5}$, and  its non-linear nature rules out the Ising nature of the spins within the clusters. We can conclude that the interaction between spins in the GP regime are Heisenberg-like. The TRM and IRM measurements provide strong evidence in favour of the presence of FM clusters  within the PM matrix~\cite{GPmemory1,nair}.  

\par
Fig.~\ref{cp} shows the zero-field heat capacity ($C_p$) data measured between 2-300 K, where  $C_p$ decreases monotonically with decreasing $T$. $C_p$ tends to saturate above 250 K and its value at 300 K is found to be 105 J mol$^{-1}$K$^{-1}$. This value is marginally higher than the  classical Dulong-Petit value of 12$R$/mol = 99.8 J mol$^{-1}$K$^{-1}$ ($R$ is the universal gas constant). Such deviation in $C_p(T)$ is a signature of phonon anharmonicity in presence of  disorder~\cite{anh1,anh2}. Notably, $C_p$ does not show any $\lambda$-like anomaly at around $T_N$.  However, $C_p/T$ vs $T^2$ data show a shallow minimum below $T_N$ (see inset), which possibly related to the AFM-like order in the system. The $C_p/T$ vs $T^2$ varies linearly between $T$ = 12 and 20 K, and the Sommerfeld coefficient $\gamma$ obtained from this linear region is found to be 26 mJ mol$^{-1}$K$^{-2}$~\cite{Cp}. This value is twice as high than the full Heusler Fe$_2$TiSn~\cite{fe2tisn}. The absence of $\lambda$-like anomaly in the heat-capacity data is not uncommon among Heusler family of compounds~\cite{mondal,bhobe}.  
 
\par
 In many instances, spin glass are found to show a broad hump like feature in the heat capacity ($C_p$ versus $T$). In our data, we rather observe a sharp rise in the $Cp/T$ vs $T^2$ plot, which is not a standard feature of the glassy magnetic phase~\cite{bhobe}. Such low-$T$ upturn below $T_N$ can be taken as an evidence for the onset of bulk magnetic ordering in the sample. Due to the  presence of Griffiths-like short range correlation, a significant amount of magnetic entropy is already released when the sample is cooled down to $T_N$ from 300 K. Such pre-existing magnetic correlation will effectively weaken the  signature in $C_p$ at $T_N$~\cite{sampath,lu}. 

\begin{figure}
	\centering
	\includegraphics[width = 8 cm]{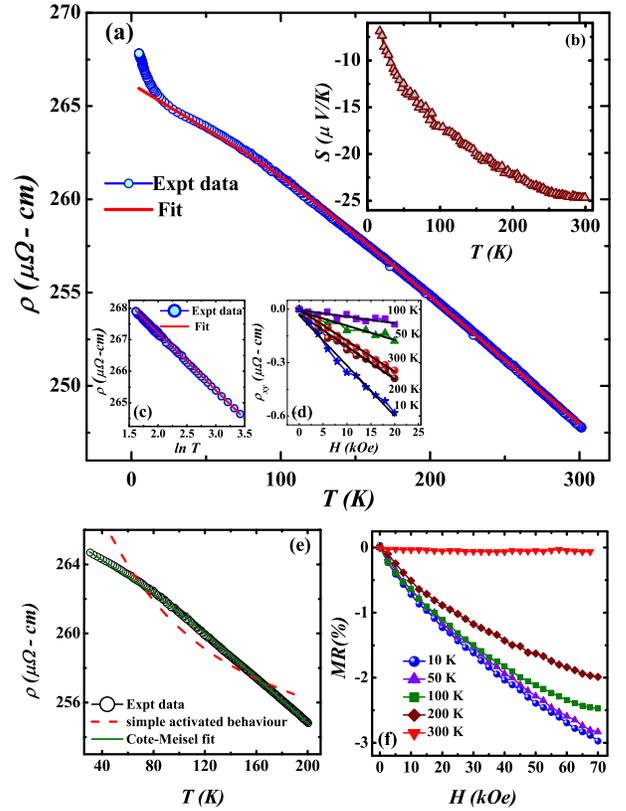}
	\caption {(a) Resistivity as a function of temperature. (b) shows the $T$ variation of thermopower. (c) presents the $\rho$ vs $\ln T$ data in the range 5-50 K (d) presents $\rho_{xy}$ vs $H$ at different constant temperatures. (e)  $\rho(T)$ data along with the fitting (black solid line) using the model proposed by Cote and Meisel(see eqn~\ref{ziman1} in the text). The red dashed line shows the common activated behaviour (convex upward) (f) Magnetoresistance measured at different temperatures.}   
	\label{res}
\end{figure}

\subsection{Electrical Resistivity and Thermopower}

The $\rho(T)$  behaviour of NiFeTiSn [see fig.~\ref{res}(a)] is found to have a  negative ($d\rho/dT <$ 0) temperature coefficient of resistivity (TCR). $\rho(T)$ is also characterized by a sharp rise below 12 K. Several quaternary and half Heusler compounds, such as CoFeCrGa~\cite{cofecrga}, CoFeMnSi, CrVTiAl~\cite{venka}, YPdSb, LuNiSb~\cite{ypdsb}, show negative TCR over a wide $T$-range. Interestingly, the nature of the  $\rho(T)$ data above 20 K in NiFeTiSn  is found to be rather unusual as compared to the common semiconducting materials. In a semiconductor, the carriers are thermally excited across an energy gap, and one generally finds an exponential (or sometimes stretched exponential) type of $\rho(T)$ data where the $\rho$ versus $T$ plot is convex upward. This has been exemplified by an exponential resistivity behaviour (red dashed line) in fig.~\ref{res}(e). On the contrary, NiFeTiSn shows a concave upward nature, where the curvature is opposite to that of common semiconductors.      

\par
The negative TCR in disordered metallic systems is often considered in the framework  of generalized Faber-Ziman theory for liquid metals and further extended for amorphous and crystalline solids by Cote and Meisel~\cite{cotemeisel,cote,faber,mizutani1,ypdsb,harris}. In this model, the resistivity of highly resistive metals can be expressed as

\begin{equation}
\rho(T) = \exp(-2\overline {W}(T))(\rho_0 + \Delta_{ep}) ~~~~~~~~~~~~~~~~~~~~~~~~~~\\
\label{ziman1}
\end{equation}
where,
\begin{equation}
\overline{W(T)} = \frac{3h^{2}<k^{2}>T^{2}}{8\pi^{2}m k_{B}\theta_D^{3}} \int_{0}^{\frac{\theta_D}{T}} \left [\frac{1}{{\rm e}^z - 1} + \frac{1}{2}\right ]zdz~~~~~~~~
\label{ziman2}
\end{equation}

Here, $\rho_0$ is the residual resistivity due to the scattering of electrons with static defects, and $\Delta_{ep}$ is the inelastic electron-phonon interaction term. 2$\overline {W}(T)$ is the Debye-Waller exponent averaged over all the scattering vectors $k$ of the conduction electrons.  In eqn.~\ref{ziman2}, $m$ stands for the ionic mass, while $h$, $k_B$ and $\theta_D$ are respectively Planck constant, Boltzmann constant and Debye temperature. For  disordered metals with high value of $\rho_0$ , $\Delta_{ep}$ is negligible as compared to  $\rho_0$. Since 2$\overline {W}(T)$ increases with increasing $T$,  eqn.~\ref{ziman1} will lead to a negative TCR. We have fitted eqn.~\ref{ziman1}  to the $\rho(T)$ data with an approximation that $\Delta_{ep}$ is negligible as compared to $\rho_0$ for NiFeTiSn. The fitting [fig~\ref{res}(a)] converges well between 70 to 300 K and we obtain $\rho_0$ = 270 $\mu\Omega$-cm, $<k^{2}>$ = 5.88$\times {\rm10^{21}}$ m$^{-2}$ and $\theta_D$ = 327 K.  

\par
To explain the low-$T$ upturn in  $\rho(T)$,  we have plotted the data (5-50 K)  as a function of $\ln T$ [ fig.~\ref{res} (c)]. A linear variation indicates the upturn to be logarithmic in nature. Such upturn may arise due to weak localization or Kondo effect~\cite{kondo1}.
	
\par
 Notably, NiFeTiSn presents a small ($\sim -$3\%) negative MR at $T$ = 10 K as shown in fig.~\ref{res}(f). With increasing $T$, the magnitude of the  MR decreases and it is about $-$2\% at 200 K. However, the MR turns negligible at 300 K. The finite negative MR at 200 K, which is well above the magnetic ordering temperature, is in line with the short range magnetic correlations.

\par
$T$ dependence of Seebeck coefficient ($S$) was measured for  NiFeTiSn  and shown fig.~\ref{res}(b). The negative value of $S$ indicates that  electron is the dominant carrier. $S$ increases with $T$ following a quasilinear  behaviour up to room temperature. The maximum value of $|S|$ is found to be around  25 $\mu$V/K at room temperature. The absolute value of $S$ at 300 K for NiFeTiSn is comparable to that of potential thermoelectric Heusler alloys such as Fe$_2$VGa and Ru$_2$NbGa ($\sim$ 25 $\mu$V/K)~\cite{lue}.  NiFeTiSn has appreciable value of thermoelectric power factor PF = $S^2/\rho$ of  2.5 $\mu$W.cm$^{-1}$.K$^{-2}$ at 300 K. This value of PF  is comparable to good thermoelectric materials such as  semimetallic Ru$_2$NbGa  and Ru$_2$NbAl~\cite{lue1,rnbga,mondal}.  We have also measured the transverse Hall resistivity ($\rho_{xy}$) and its variation with  $H$ is shown in  fig.~\ref{res}(d).  We have calculated the  values of carrier concentration ($n$) and Hall mobility ($\mu_H$) at $T$ = 10, 300 K, and they are found to be $n$ = 2.3$\times$10$^{21}$, 2.2$\times$10$^{21}$cm$^{-3}$ and 10.2, 11.5 cm$^2$/V-s respectively. Negative Hall coefficient also supports the result obtained from $S(T)$ measurement, {\it i.e.}, electrons are the dominant carriers in NiFeTiSn. The $\mu_H$ obtained for NiFeTiSn is comparable to other reported Heusler compounds~\cite{mob1,mob2}.

\section{Discussion}

There are only limited number of reports for Griffiths-like phase in Heusler alloys and compounds. They include systems such as Fe$_2$VGa, Fe$_2$VAl and Fe$_2$CrAl-based alloys. The present NiFeTiSn is a quaternary Heusler compound, and to our knowledge, FeMnCrAl is the only other compound showing GP in this series~\cite{kavita}. However, unlike NiFeTiSn, FeMnCrAl has a long range FM ground state. The cusp like feature around 7 K both in the ZFC and FC magnetisation data indicates that the ground state is AFM-like in NiFeTiSn [fig.~\ref{mag} (a)].
\par
 So far, most of the experimentally reported intermetallic GP compounds order ferromagnetically at low temperature~\cite{kavita,GP3,pramanik}, and AFM GP compounds are relatively lesser in number. The signature of GP above $T_N$ is quite convincingly established from our study. The archetypal signature of GP is evident from the $\chi^{-1}$ versus $T$ data, where we observe a strong downturn from linear Curie-Weiss law below $T_{GP}$ ($\approx$ 275 K). We also observe a power law variation of inverse susceptibility ($\chi^{-1} \sim T^{1-\lambda}$), which is a common protocol to establish a GP. The magnetisation of NiFeTiSn also obeys the scaling law, $\ln{M} \sim -\frac{T}{H}$ roughly in the temperature range between $T_N$ and $T_{GP}$.
\par    
Microscopically a GP is characterized by spatially isolated non-interacting FM clusters in a  paramagnetic matrix. Therefore, it is pertinent to show that our compound is devoid of any long-range order between $T_N$ and $T_{GP}$. If we extrapolate the high $H/M$ Arrot plot data (above $T_N$) towards the  origin [see fig.~\ref{arrot}], they strike the ordinate at negative values of $M^2$, indicating the lack of spontaneous magnetisation. Further proof opposing the long range ordered phase between $T_N$ and $T_{GP}$ is obtained from our M\"ossbauer spectra. The distinct singlet line obtained both at high-$T$ (300 K) and low-$T$ (80 K) confirms the lack of a long range magnetic order in these $T$ range for NiFeTiSn.  The M\"ossbauer line recorded at 80 K ($\gg T_N$) is found to be significantly wider than that of 300 K data. This increase in the width  possibly indicates the growth of the short range FM clusters  with decrease in $T$. Similar observation were reported in the Heusler compounds Fe$_2$VGa and Fe$_2$VAl~\cite{mossbauer2,mossbauer3} and Eu-doped Ca$_3$Co$_2$O$_6$~\cite{paulose}.

\par
The unexpected magnetic anomaly observed close to room temperature could also be associated to the existence of ferromagnetic impurity phase with a $T_C$ close to room temperature. However, such possibility can be convincingly ruled out in light of the following evidences: (A) Lack of spontaneous magnetisation (for $T > T_N$) indicates the absence of long-range ferromagnetic ordering and  thus excludes the possibility of the existence of a FM impurity phase. (B) The M\"ossbauer spectra have singlet line at 300 K and 80 K. The line is sharp and symmetric. Had there been any ferromagnetic impurity or ordering, one should see a splitting or broadening of the line. $^{\rm 57}$Fe M\"{o}ssbauer spectra is quite sensitive to any ordered impurity phase. For example, even 1\% Fe (non-enriched) doping in BaTiO$_3$ can produce significant splitting of the singlet line~\cite{batio3}. However, in case of NiFeTiSn, we find the M\"{o}ssbauer line which can be well fitted by a single Lorentzian function [see fig.~\ref{xrd}(b)]. This certainly signifies that the sample does not order magnetically at least down to 80 K within the accuracy of the M\"{o}ssbauer technique. The lack of magnetic order is the most important characteristics of a Griffiths-like phase and the present data  supports the presence of GP in the studied compound. (C) Careful observation and analysis of x-ray powder diffraction data for NiFeTiSn reveals the absence of any major impurity phase (within the accuracy limit of the experimental technique). All the peaks observed can be indexed to the cubic LiMgPdSn type Heusler structure~\cite{marcano}. (D) The nonexponential decay of $M$ with time above $T_N$ indicates isolated non-frustrated spin clusters~\cite{GPmemory1} and it rules out long-range ordering of the FM impurity phase. Above all, $\chi$ and $M$ follow the appropriate scaling laws of a GP as opposed to the behaviours of an FM state.

\par
Magnetic relaxations measured in the GP-regime of the sample [see fig~\ref{relax}(a)] does not obey a simple  exponential law. Instead, we observe $M(t)$ to follow  $\exp[-\sqrt{(\frac{t}{\tau})}]$ type of variation with time. Theoretical arguments indicate that such form of relaxation is a signature of Heisenberg type spin system within the FM clusters of the GP~\cite{griffiths2,griffiths3}. As an example, REBaCo$_4$O$_7$ (RE = rare-earth) compounds show similar relaxation behavior, and the Heisenberg nature of the spin system is further verified by means of neutron scattering experiments~\cite{manuel}.

\par
NiFeTiSn has a simple cubic structure at room temperature  and it does not show any change in its lattice symmetry as a function of temperature~\cite{maple}. The full Heusler compound Fe$_2$TiSn is an important example where ferromagnetism originates from antisite disorder. Our PXRD analysis indicates about 17\% antisite disorder between Ti and Fe sites. Such disorder can create isolated Fe-rich magnetic clusters leading to Griffiths-like phase. Interestingly, NiFeTiSn shows  GP from 280 K, which is very close to the Curie temperature of Fe$_2$TiSn. DFT based calculations indicate small value of moment at the Ni site of the quaternary compounds NiFeTiZ (Z = P, As, Si, Ge)~\cite{karimian}. As compared to Fe$_2$TiSn, one Fe atom is replaced by Ni in NiFeTiSn, and this dilution of Fe atoms might be responsible for the lack of long range FM order developing around 260-280 K. Instead, isolated FM clusters are formed in the latter compound leading to Griffiths-like phase. It is worth noting  that GP in Fe$_2$VGa and some  of its dopants is also related to antisite disorder between Fe and V sites~\cite{fevga}.

\par
In the original conjecture laid down by Griffiths, the short range FM clusters in the GP can eventually lead to a long-range ordered  FM ground state on lowering the temperature. However, the situation can be different if the ground state attains antiferromagnetism. There are numerous examples in the literature on  such antiferromagnetic Griffiths-like phase, and the present  NiFeTiSn is found to be a new addition to the list. The most pertinent question regarding the  AFM GP is the nature of the magnetic ground state. In the literature, there are reports where the short range FM clusters continue to coexist with the long-range state at low temperature (example, Ca$_3$CoMnO$_6$~\cite{cacomno}). Such coexistence can give rise to spin glass or cluster glass like phase arising from the competition between FM clusters and the AFM phase. For example, glassy magnetic phase is observed below $T_N$ in the double perovskites  La$_2$FeMnO$_6$ and Pr$_{2-x}$Sr$_x$CoMnO$_6$~\cite{LaFeMnO,PrCoMnO}. Many of these compounds also show exchange bias effect at low temperature, which is an evidence for magnetic inhomogeneity. On the contrary, electron spin resonance experiment  categorically ruled out the presence of short range FM phase below $T_N$ in the manganite Eu$_{0.6}$La$_{0.4-x}$Sr$_x$MnO$_3$~\cite{eremina}. 

\par

In a magnetic solid, GP can have different origins, such as quenched disorder (in manganites), phase separation, micro-twinning, competing intra and inter-layer magnetic interactions (in a quasi-two dimensional system), cluster formation in nano-sized materials and so on. Among intermetallics, GdSn$_2$ based alloys are  also claimed to have GP along with AFM ground state. On the basis of x-ray powder diffraction coupled with the spin-polarized density functional calculations, Ghosh $et. al.$ established a direct structural connection to GP in intermetallic GdM$_x$Sn$_{2-\delta}$ (M = Co, Ni, Cu; 0 $< x <$ 1, $\delta$ = 0-0.22)~\cite{gdsn2,ghosh}. Griffiths-like phases are also reported in the magnetically disordered rare-earth based intermetallic compounds Tb$_5$Si$_2$Ge$_2$ and RE$_5$Si$_x$Ge$_{4-x}$ series~\cite{gdsn2,GP3,GdGe}. In this case, the occurrence of GP is microscopically associated  with competing intra- and interlayer exchange interactions, due to their complex structure. In contrast, NiFeTiSn belongs to cubic Heusler series  containing 3$d$ transition metals and $sp$ element only. The cubic structure is retained down to low temperature~\cite{maple} and it  is pretty much three dimensional in nature without having any chance for quasi-low dimensional magnetic correlation. Unlike manganites and cobaltites, the system is devoid of electronic phases separation. The phase separation due to structural transition, observed in Heusler based shape memory alloys, is also absent here as the system retains the cubic Heusler structure down to low temperature. Hence, it can be assumed that the structural correlation is not the reason for the formation of GP in the cubic NiFeTiSn. Rather, it is more likely that antisite disorder is the key driving force for the observed GP-like anomaly. 

\par         
In case of NiFeTiSn, we failed to observe any signature of FCFS memory neither below nor above $T_N$, and a spin glass/superparamagnetic phase can be ruled out for the compound. The compound does not show any signature of exchange bias at 2 K (see inset of fig~\ref{mag}(d)), which also demonstrates that there is no major magnetic inhomogeneity below $T_N$. The $\chi^{-1}$ versus $T$ data [fig.~\ref{invchi} (b)], measured at different $H$, converge at $T_N$. Had there been any FM cluster present below $T_N$, one would expect $\chi^{-1}$ to follow different paths depending upon $H$. These observations, apparently points that the GP ceases to exist below $T_N$ in NiFeTiSn. 
\par   
Nevertheless, the ground state of NiFeTiSn is not purely AFM as evident from the magnetic isotherms depicted in fig.~\ref{mag}(d). The $M-H$ curves below 50 K show `S'-like nature, and a finite coercivity of 200 Oe and a remanent magnetisation of 48 emu/mol are observed  at 2 K [inset of fig.~\ref{mag} (d)]. The finite coercivity undoubtedly indicates the presence of an ordered FM component below $T_N$, and the resultant magnetic structure is likely to be canted AFM type. The short range FM correlation present in the sample in the form of GP might be instrumental for spin canting when the sample attains a long-range order below $T_N$. It is to be noted that the value of spontaneous magnetisation obtained from the Arrot plot is quite small, which corresponds to the  weak ferromagnetism  below $T_N$ due to spin canting~\cite{regnat,chadov}. However, we also have to rule out the possibility of the sample being a simple ferrimagnet. In case of a ferrimagnet, there have to be two antiparallel sublattices with different magnetic moment values. Consequently, the net moment will be non-zero. For the present NiFeTiSn, there can be Ni and Fe sublattices.  The band structure calculations of similar compounds such as NiFeTiZ (Z = P, Si, Ge, As) shows that the moment at the Fe site ($m_{{\rm Fe}}$) is much larger than that of the   Ni site ($m_{{\rm Ni}}$)~\cite{karimian}. Therefore the net moment ($m = m_{{\rm Fe}} - m_{{\rm Ni}}$) will be significant. Since our data indicates very small value of $M_S$, the scenario for  a ferrimagnetic state is improbable.    

\par
 NiFeTiSn shows a negative thermal variation of $\rho$, but does not resemble a general activated behavior found in semiconductors. Rather $\rho(T)$ shows an unusual nature where the curve is {\it concave downwards}. Fe$_2$TiSn, on the other hand, shows metallic behaviour  below about $T_C$, and shows negative temperature coefficient behaviour of resistivity ($d\rho(T)/dT <$ 0) in the PM region~\cite{fe2tisnconcave}. The metallicity in Fe$_2$TiSn is believed to be connected to the long range FM order. Since NiFeTiSn does not order ferromagnetically at high temperature, it retains its negative temperature coefficient of resistivity down to low temperatures. The  $\rho(T)$ data between $T_{GP}$ and $T_N$ can be fitted with Cote-Meisel model~\cite{cotemeisel} for disordered systems. The scattering of conduction electrons with static site disorders and/or  isolated magnetic clusters is possibly responsible for the rather unusual $\rho(T)$ variation in this temperature window.  Interestingly, the Seebeck coefficient of NiFeTiSn  is found to be negative as compared to the positive value in Fe$_2$TiSn, although the magnitudes of thermopower are comparable~\cite{mob2}. As shown in fig~\ref{res}(b), the magnitude of $S$ increases with increasing temperature following a quasilinear behavior up to room temperature. For heavily doped semiconductors (parabolic band, energy-independent scattering approximation) with degenerate character of electrons/holes, the magnitude of $S$ generally increases monotonically with increasing $T$.~\cite{mob2,snyder,carr}.
   
\par
In conclusion, we report an extensive investigation on the  cubic  Heusler compound NiFeTiSn by means of magnetic, M\"ossbauer, electrical and thermal transport properties measurements. The work establishes a Griffiths-like phase that exists over a wide range of temperature (15-275 K). The ground state of the compound is found to be canted AFM type in nature, and our work indicates that the FM clusters due to GP disappear with the advent of this canted AFM state below $T_N$.  

\section*{Acknowledgment}
SC wishes to thank UGC, India for his research fellowship. The work is supported by the grant from SERB, India with grant number EMR/008063.


\end{document}